\documentclass[twocolumn]{aastex61}

\usepackage{hyperref}
\usepackage{amsmath}

\received{Febraury 9, 2018}
\revised{December 21, 2018}
\accepted{January 11, 2019}
\submitjournal{ApJ}

\shorttitle{First Spectral Analysis of a Solar Plasma Eruption Using ALMA}
\shortauthors{Rodger et al.}

\begin{document}

\title{First Spectral Analysis of a Solar Plasma Eruption Using ALMA}

\correspondingauthor{Andrew Rodger}
\email{a.rodger.1@research.gla.ac.uk}

\author{Andrew S. Rodger}
\affil{SUPA, School of Physics and Astronomy, University of Glasgow, Glasgow G12 8QQ, United Kingdom}

\author{Nicolas Labrosse}
\affiliation{SUPA, School of Physics and Astronomy, University of Glasgow, Glasgow G12 8QQ, United Kingdom}

\author{Sven Wedemeyer}
\affiliation{Rosseland Centre for Solar Physics, University of Oslo, P.O. Box 1029 Blindern, N-0135, Norway}
\affiliation{Institute of Theoretical Astrophysics, University of Oslo, P.O. Box 1029 Blindern, N-0135, Norway}

\author{Mikolaj Szydlarski}
\affiliation{Rosseland Centre for Solar Physics, University of Oslo, P.O. Box 1029 Blindern, N-0135, Norway}
\affiliation{Institute of Theoretical Astrophysics, University of Oslo, P.O. Box 1029 Blindern, N-0135, Norway}

\author{Paulo J.A. Sim\~oes}
\affiliation{SUPA, School of Physics and Astronomy, University of Glasgow, Glasgow G12 8QQ, United Kingdom}
\affiliation{Centro de R\'adio Astronomia e Astrof\'isica Mackenzie, CRAAM, Universidade Presbiteriana Mackenzie, S\~ao Paulo, SP 01302-907, Brazil}

\author{Lyndsay Fletcher}
\affiliation{SUPA, School of Physics and Astronomy, University of Glasgow, Glasgow G12 8QQ, United Kingdom}



\begin{abstract}

The aim of this study is to demonstrate how the logarithmic millimeter continuum gradient observed using the Atacama Large Millimeter/submillimeter Array (ALMA) may be used to estimate optical thickness in the solar atmosphere. 
We discuss how using multi-wavelength millimeter measurements can refine plasma analysis through knowledge of the absorption mechanisms.
Here we use sub-band observations from the publicly available science verification (SV) data, whilst our methodology will also be applicable to regular ALMA data. 
The spectral resolving capacity of ALMA SV data is tested using the enhancement coincident with an X-ray Bright Point (XBP) and from a plasmoid ejection event near active region NOAA12470 observed in Band 3 (84--116~GHz) on 17/12/2015. 
We compute the interferometric brightness temperature light-curve for  both features at each of the four constituent sub-bands to find the logarithmic millimetre spectrum. 
We compared the observed logarithmic spectral gradient with the derived relationship with optical thickness for an isothermal plasma to estimate the structure's optical thicknesses. 
We conclude, within 90\% confidence, that the stationary enhancement has an optical thickness between 0.02 $\leq \tau \leq$ 2.78, and that the moving enhancement has 0.11 $\leq \tau \leq$ 2.78, thus both lie near to the transition between optically thin and thick plasma at 100~GHz. 
From these estimates, isothermal plasmas with typical Band 3 background brightness temperatures would be expected to have electron temperatures of $\sim$7370 -- 15300~K for the stationary enhancement and between $\sim$7440 -- 9560~K for the moving enhancement, thus demonstrating the benefit of sub-band ALMA spectral analysis.

\end{abstract}

\keywords{Sun: Chromosphere -- millimeter continuum -- radiative transfer}



\section{Introduction} \label{sec:intro}

The Atacama Large Millimetre/sub-millimetre Array (ALMA) has the potential to be a revolutionary tool for modern solar physics providing high resolution interferometric measurement in a previously less-explored spectral window. 
The quiet solar chromosphere emits mm/sub-mm radiation, in the Rayleigh-Jeans limit, predominantly through thermal bremsstrahlung which is a local thermodynamic equilibrium (LTE) emission mechanism.
Therefore, a brightness temperature measurement from optically thick source material will be highly representative of the electron temperature of the region in which the emission was formed \citep{2016SSRv..200....1W,2017SoPh..292..130R}.
Until ALMA however, mm/sub-mm imaging has lacked sufficiently high resolution to allow for in-depth analysis at the small scales critical for understanding many solar atmospheric processes. 

The first ALMA solar observing cycle (cycle 4) was conducted in 2016--2017. 
In cycle 4 the ALMA modes and capabilities available for solar physics were Bands 3 (84--116~GHz) and 6 (211--275~GHz) using the most compact-array configurations (maximum baselines $< 500$m) at an imaging cadence of $\sim$2s. 
\cite{2017SoPh..292...87S} give an account of the ALMA solar science verification (SV) efforts including descriptions of the required Mixer-Detuning method of receiver gain reduction and calibration processes for ALMA solar data. 
They also discuss how to estimate the noise level for interferometric images using the difference between cross-correlated orthogonal linear polarization measurements.
Absolute brightness temperature measurements from ALMA require the interferometric images to be ``feathered'' with measurements taken using a set of up to four separate total-power (TP) antennas. 
\cite{2017SoPh..292...88W} provide a description of the Fast-Scanning Single-Dish Mapping technique employed by ALMA's TP antennae. 
Other publications using the SV data include \cite{2017A&A...605A..78A} who examine the centre-to-limb variation observed in the mm/sub-mm domain, \cite{2017ApJ...845L..19B} who compare mm and Mg~II ultra-violet emission, and \cite{2017ApJ..841L..20I} who report a brightness enhancement at 3~mm in a sunspot umbra.

In this article we present the diagnostic capability of ALMA with a focus on Band~3 using measurements at each of its four constituent sub-bands, also known as \emph{spectral windows} (or spw). 
The method is applicable to other bands available to solar observations.
Through the measurement of the brightness temperature at several frequencies within one ALMA Band, it is possible to construct a millimetre continuum spectrum providing more constraints for the emission mechanism from a region and to refine the diagnostic of the plasma conditions. 
To do this we use the relation between optical thickness of emitting material and logarithmic spectral gradient which is discussed for an off-limb case in \cite{Rodger&Labrosse2018}.
We demonstrate this using ALMA Band 3 observation of a plasmoid ejection from active region NOAA12470 from the 17th of December 2015.
This provides an interesting case to study due to the enhancement in brightness temperature caused by the plasmoid observed.
This event has been analysed by \cite{2017ApJ...841L...5S} who set limits on the possible density and thermal structure of the plasmoid using the brightness temperature integrated across Band 3, observations at EUV wavelengths from \emph{SDO}/AIA and soft X-rays using \emph{Hinode}/XRT. 
They calculate the average enhancement observed in the plasmoid at ALMA Band 3 (100~GHZ) and the 171, 192 and 211~\AA\ AIA Bands.
From these they obtain the required density/temperature curves for formation aiming to find areas of cross-over between the ALMA and AIA bands. 
They conclude that the plasmoid consists of an isothermal ~$10^5$~K plasma that is optically thin at 100~GHz, or a multi-thermal plasmoid with a cool ~$10^4$~K core and a hot EUV emitting envelope. 

Section~\ref{sec:theory} briefly presents how the millimetre continuum is formed and how it may be used to distinguish between differing optical thickness and thermal structure models. 
In Section~\ref{sec:obs} we describe the data used and the methods for image synthesis and calculation of plasmoid brightness temperature enhancement. 
We present our results in Section~\ref{sec:results} and a discussion of the results is given in Section~\ref{sec:discussion}. 
Conclusions are given in Section~\ref{sec:conc}.

\section{Formation of the mm continuum in the quiet Sun} \label{sec:theory}

The primary emission mechanism for millimetre radiation from the quiet chromosphere is thermal bremsstrahlung.
This process is purely collisional allowing the radiation to be described simply by local thermodynamic equilibrium processes.  
Millimetre radiation being in the Rayleigh-Jeans limit means that for an optically thick plasma the observed brightness temperature represents an accurate measurement of the electron temperature in the region of continuum formation. 
However, an optically thin plasma will have a brightness temperature lower than the electron temperature.
Here we briefly recall the main question that arises when interpreting the observed brightness temperature, namely  whether the plasma is isothermal or not.

\subsection{Isothermal plasma}

The simplest case is when the plasma is isothermal. If the plasma is optically thick over a range of wavelengths, the brightness temperature spectrum would be flat over that wavelength range, at the electron temperature value.
If the plasma is optically thin however, the brightness temperature spectrum would not be flat and would vary according to the optical thickness $\tau(\nu)$ at each wavelength. 
The following equations show the frequency dependent brightness temperature for an optically thin, isothermal plasma calculated using the absorption coefficient as described by \cite{1985ARA&A..23..169D};

\begin{equation}\label{eq:opthin_isoT}
T_{\mathrm{B}}(\nu) = 1.77 \times 10^{-2} \times \langle \mathrm{EM} \rangle \frac{ g_{\rm{ff}}(T,\nu) }{\nu^{2}T^{\frac{1}{2}}} \ , 
\end{equation}
where $T$ is the electron temperature, $\nu$ is the frequency of observation, and $g_{\rm{ff}}(T,\nu)$ is the free-free Gaunt factor interpolated from the table of numerically calculated values given by \cite{2014MNRAS.444..420V} \citep{1970A&A..9..312G,2017A&A..605A.125S}. 
The average column emission measure ($\langle EM \rangle$) for a layer of thickness $L$ is defined as in \cite{2017SoPh..292..130R}:
\begin{equation}\label{eq:EM}
\langle \mathrm{EM} \rangle= \langle n_{\mathrm{e}} \sum_j Z_j n_j \rangle L . 
\end{equation}
$n_{\mathrm{e}}$ is the electron density with $Z_j$ and $n_j$ being the charge and density of ion species $j$, respectively. 
From Equation~\ref{eq:opthin_isoT} it can be seen that, for an optically thin, isothermal source, the mm continuum behaves as $g_{\rm{ff}}(T,\nu) \times \nu^{-2}$.  

As shown in \cite{Rodger&Labrosse2018} the spectral gradient of the logarithmic millimetre brightness temperature spectrum for an isothermal source can be described as: 
\begin{equation}\label{eq:spectral_gradient_off_limb}
\frac{\mathrm{d\,log}(T_{\mathrm{B}})}{\mathrm{d\,log}(\nu)} = \alpha \frac{-2\tau_{\nu}}{\mathrm{e}^{\tau_{\nu}}-1},
\end{equation}
where $\alpha$ is a correcting factor caused by a non-zero rate of change of gaunt factor, $g'_{\mathrm{ff}}$, with frequency, over the frequency band. 
$\alpha$ is defined by:
\begin{equation}\label{eq:alpha}
\alpha = 1 - \frac{\nu g'_{\mathrm{ff}}}{2{g_{\mathrm{ff}}}} .
\end{equation}
It was found in \cite{Rodger&Labrosse2018} that equation~\ref{eq:spectral_gradient_off_limb} displays a clear relationship between logarithmic spectral gradient and optical thickness regime for any isothermal plasma, provided that suitable bounds are set on the value for the $\alpha$ factor. 
The diagnostic may be used to find (a) whether the plasma is in the fully optically thin regime ($\tau \lesssim 10^{-1}$), (b) the optical thickness of the plasma if it lies within the transition between optically thin and thick plasma ($10^{-1} \lesssim \tau \lesssim 10^{1}$), or (c) whether the plasma is in the fully optically thick regime ($\tau \gtrsim 10^{1}$).

\subsection{Multi-thermal plasma}

A multi-thermal case can be significantly more complex. 
For an optically thick multi-thermal plasma the brightness temperature at a given wavelength will be representative of the temperature around the region of formation of the continuum at that wavelength \citep{2017SoPh..292..130R,2017A&A..605A.125S}. 
As the optical thickness decreases with increasing frequency, the mm continuum formation region will be deeper into the observed structure. 
The brightness temperature spectrum in this case will not be flat as in the isothermal case but will vary depending on the thermal structure of the plasma. 

If the plasma is optically thin we would expect a brightness temperature spectrum similar to Equation~\ref{eq:opthin_isoT} but with non-uniform temperature as follows:
\begin{equation}\label{eq:opthin_nonisoT}
T_{\mathrm{B}}(\nu) = 1.77 \times 10^{-2} \nu^{-2} \int n_e \sum_j Z_j n_j g_{\rm{ff}}(T,\nu) T^{-1/2} \mathrm{e}^{-t_{\nu}} \mathrm{d}l , 
\end{equation}
where $l$ is the path along the line of sight and $t_{\nu}$ is the optical thickness at each point along the integration path.
It has been shown in \cite{Rodger&Labrosse2018}, however, that Equation~\ref{eq:spectral_gradient_off_limb} is a suitable diagnostic relationship for an optically thin, multi-thermal plasma, despite the relation being derived from an isothermal assumption. 
However, above $\tau=1$ the logarithmic spectral gradient of the multi-thermal plasma is defined by both the optical thickness and the temperature gradient of the plasma, making Equation~\ref{eq:spectral_gradient_off_limb} less reliable there. 

If the observed mm spectral gradient is non-zero the plasma must be either optically thin, or optically thick and multi-thermal. 
A method to discern between these two scenarios may be to compare the extent of the emitting region at different wavelengths. 
If the plasma is optically thin, its physical extent will appear roughly the same at each frequency, whilst an optically thick plasma may vary in extent if the region of continuum formation varies with frequency.
To detect this requires the distance between formation regions at multiple wavelengths to be greater than the resolvable spatial scales of the observation.

\section{Observation} \label{sec:obs}

\begin{figure}
	\centering
	\includegraphics[width=\linewidth]{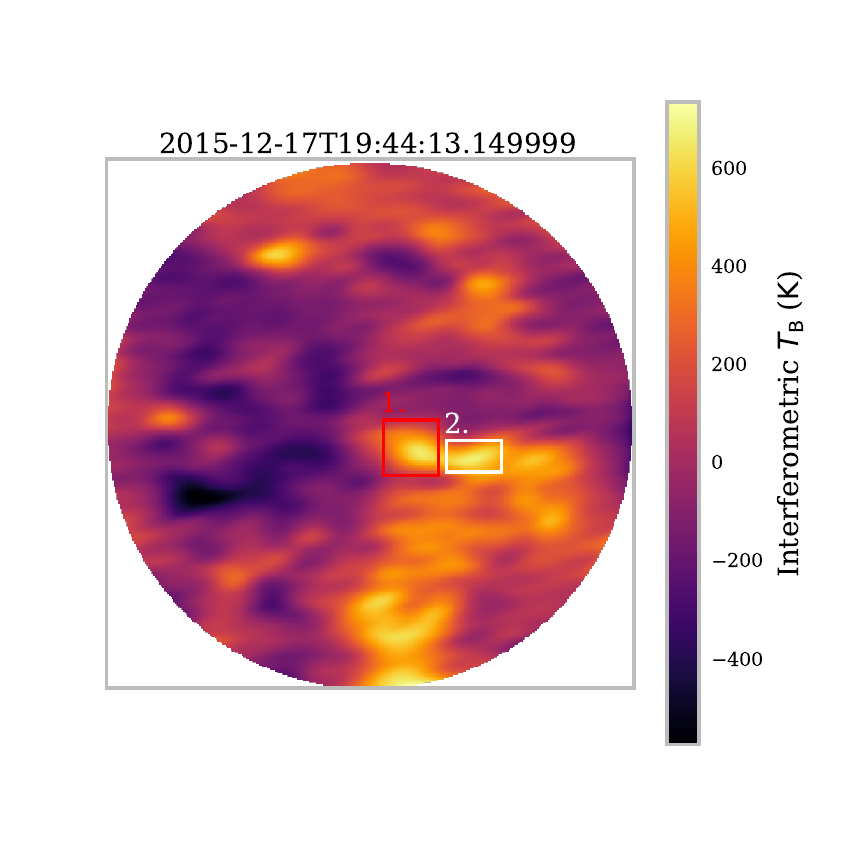}	
	\caption{Interferometric image of ALMA field of view observing active region NOAA12470 on the 17th of December 2015.
		This shows an ALMA Band 3 spectral window 0 (93~GHz) image produced in single 2s interval.
		The colourbar shows the interferometric brightness temperature in Kelvin. 
		The two boxes on the image show the location of the two regions of interest to this study. 
	}
	\label{fig:boxes}
\end{figure}
\begin{figure}
	\centering
	\includegraphics[width=\linewidth]{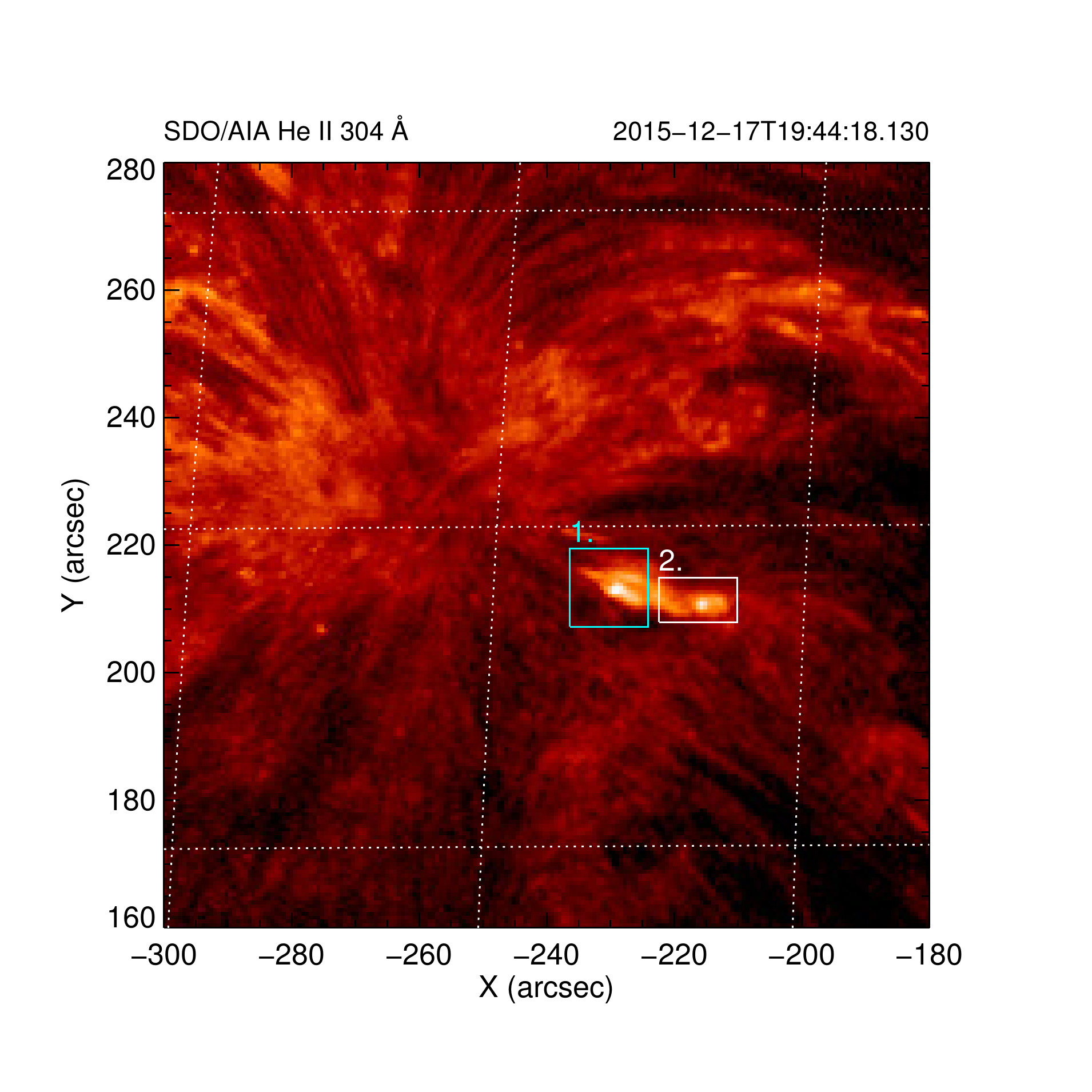}
	\caption{Context for observation and two regions of interest shown in Figure~\ref{fig:boxes} as viewed with SDO/AIA at 304~\AA.}
	\label{fig:AIA}
\end{figure}

\begin{figure}[t!]
	\centering
	
	\includegraphics[width=\linewidth]{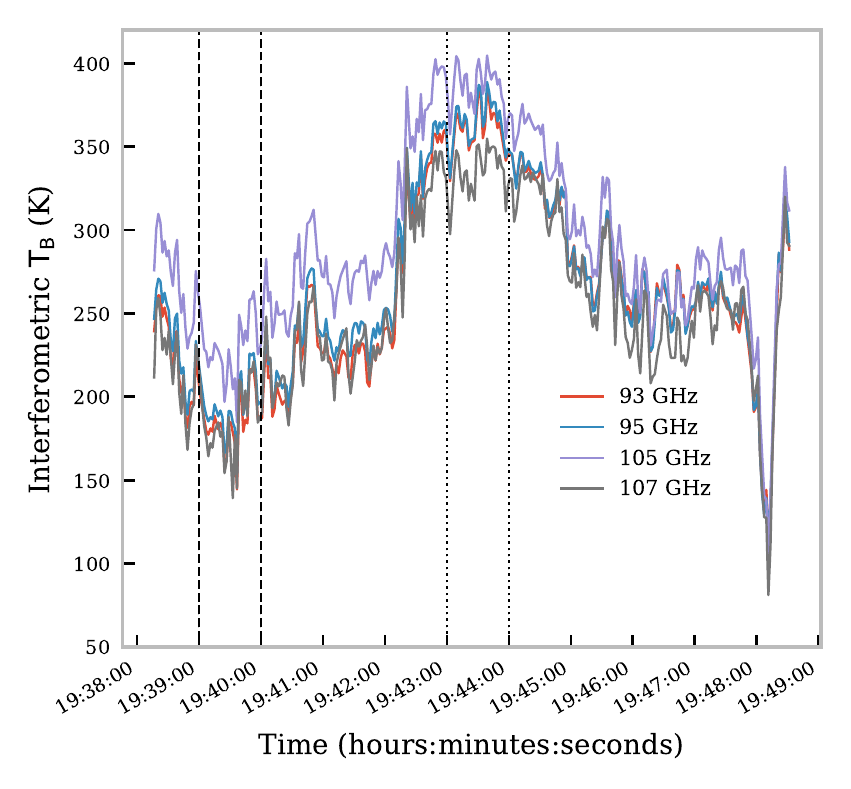}
	\caption{Lightcurve showing interferometric brightness temperature in region coincident with an XBP, box 1 in Figure~\ref{fig:boxes}, across all constituent sub-bands of ALMA Band~3. The region between the dashed lines shows the pre-enhancement background level, whilst the region between dotted lines shows plasmoid enhancement region used throughout this study.}
	\label{fig:box1lightcurve}
\end{figure}
\begin{figure}[t!]
	\centering
	\includegraphics[width=\linewidth]{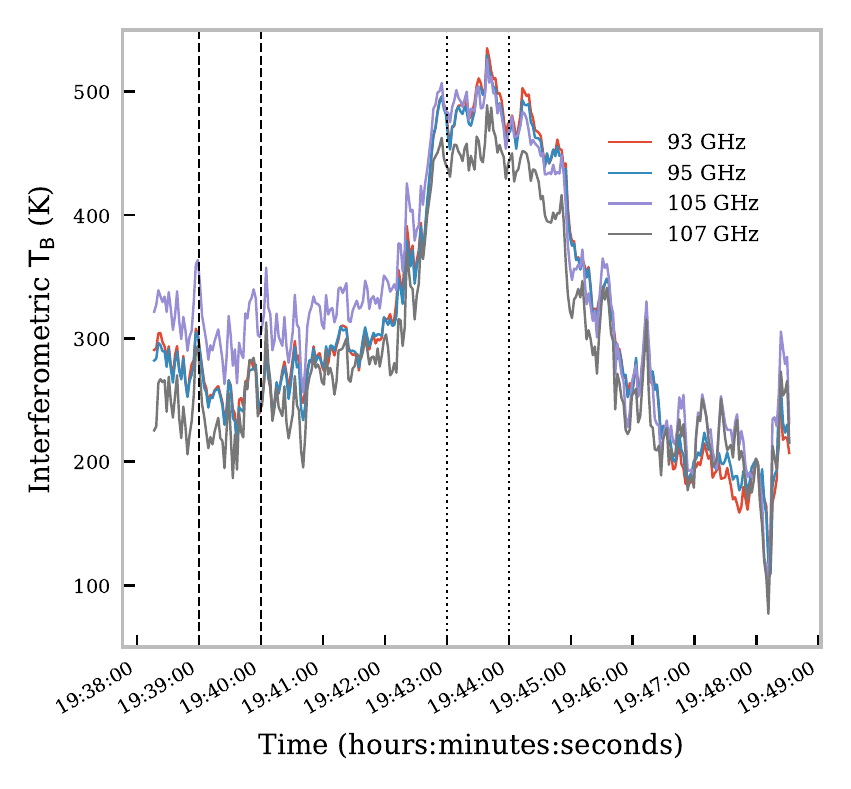}
	\caption{Same as Figure~\ref{fig:box1lightcurve}, but for box 2 in Figure~\ref{fig:boxes}.}
	\label{fig:box2lightcurve}
\end{figure}


\begin{figure*}
	\centering
	\includegraphics[width=\linewidth]{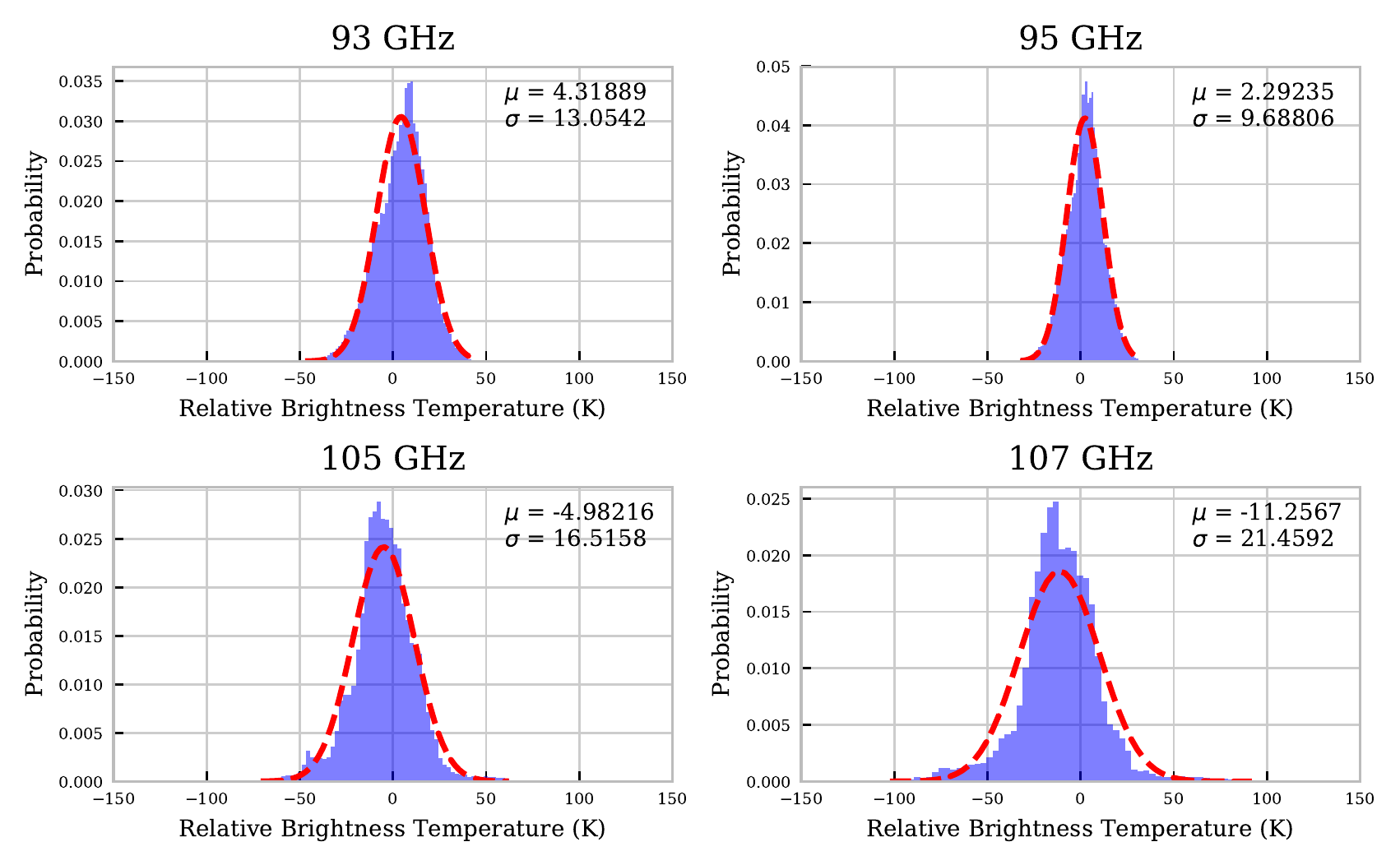}
	\caption{Noise distributions calculated using difference between XX and YY cross-correlated linear polarisation data for each sub-band of ALMA Band~3. 
		The images were synthesised over the whole bandwidth of each sub-band at a single time stamp of duration 2s.
		The histograms are fitted with a gaussian function (dashed red) with mean and standard deviation given in each panel.}
	\label{fig:spw_noise}
\end{figure*}

On the 17th of December 2015 ALMA observed a region near the large leading sunspot of active region NOAA12470 as part of a science verification campaign. 
This observation was conducted with a reduced interferometer setup of 22x12~m and 9x7~m antennae instead of up to 50x12~m and 12x7~m which will be the maximum possible array configuration available during full scientific campaigns. 
An enhancement in brightness temperature was detected near to a simultaneous X-ray bright point (XBP) observation, the brightness temperature enhancement showing the ejection of a moving bright blob of plasma or \emph{plasmoid} \citep{2017ApJ...841L...5S}. 
The observation was conducted using ALMA  Band~3 which has a central frequency of 100 GHz in the bandwidth of 84 -- 116~GHz.
Observations with ALMA at 100~GHz have a field of view of ~60\arcsec. 
The observing beam for this observation was elliptical with a semi-major axis of 6.2\arcsec and semi-minor axis of 2.3\arcsec, as the shape of the beam depends on the sky and thus changes. 
This dataset, along with other SV data sets, has been publicly released by the joint ALMA observatory\footnote{\url{https://almascience.eso.org/alma-data/science-verification}}. 

We first replicated the results of \cite{2017ApJ...841L...5S}. 
We used the scripts provided with the test data\footnote{\url{https://almascience.eso.org/almadata/sciver/2015ARBand3/}} to calibrate the data, and then synthesise each image using the full bandwidth of Band~3 at a cadence of 2~s.  
From the resulting time-series image, following \citeauthor{2017ApJ...841L...5S}, we define two boxes within the field of view  (box 1 and box 2 in Figure \ref{fig:boxes}). 
An SDO/AIA 304~\AA ~image shows the context for the observation in Figure~\ref{fig:AIA} \citep{2012SoPh..275..17L}.
Box 1 covers the region showing a stationary brightness temperature enhancement coincident with an XBP, whilst box 2 shows the region covering a moving brightness temperature enhancement from the plasmoid ejection. 
These boxes are not strictly identical to those used by \citeauthor{2017ApJ...841L...5S}, however they do share roughly the same location and extent. 
We then calculated the mean brightness temperature within each of box 1 and 2 at each time step for the duration of the observational scan containing the plasmoid ejection. 

Purely interferometric measurement can only provide the change in brightness temperature relative to some background value for the frequency-band observed.
As the field of view of this observation (60\arcsec\ for Band~3) is completely filled by the Sun, a background or quiet Sun measurement is not possible using interferometric data alone. 
ALMA can however produce true brightness temperature measurements through feathering the interferometric images with full-dish total power images.  
This will add an increased level of uncertainty to the data set \citep{2017SoPh..292...88W}, and in agreement with \citeauthor{2017ApJ...841L...5S} we have decided to focus on interferometric results solely. 

With this method we thus produce relative brightness temperature lightcurves for boxes 1 and 2.
The absolute value for the brightness temperature enhancement is calculated by taking the difference of the relative brightness temperature from the interferometric images at two separate periods within the observational scan, one representative of a quiet or background phase and the other of the enhanced phase.
This is shown in Figures~\ref{fig:box1lightcurve} and \ref{fig:box2lightcurve}.

\begin{table*}[ht!]
	\caption{Brightness temperature enhancements for boxes 1 and 2 in figure~\ref{fig:boxes} with the standard deviations of the respective normal uncertainty distributions of images at each constituent spectral window of ALMA Band 3.}\label{table:results}
	\centering
	\begin{tabular}{| c | c | c |}
		\hline
		Spectral Window (GHz) & Box 1 $E \pm \sigma(E)$ (K)& Box 2 $E \pm \sigma(E)$ (K) \\
		\hline
		Spw0 -- 93~GHz & 174 $\pm$ 6.8 & 235 $\pm$ 9.3 \\
		Spw1 -- 95~GHz & 170 $\pm$ 6.9 & 233 $\pm$ 9.2\\
		Spw2 -- 105~GHz & 156 $\pm$ 7.5 & 188 $\pm$ 9.9\\
		Spw3 -- 107~GHz & 150 $\pm$ 6.7 & 218 $\pm$ 9.3\\
		\hline
		Full Band -- 100~GHz & 159 $\pm$ 6.8 & 221 $\pm$ 9.4\\
		\hline
	\end{tabular}
\end{table*}

\subsection{Noise level calculation}\label{sec:noise}

The noise level of the synthesised images was estimated by calculating the difference image between XX and YY cross-correlations of the two orthogonal linear polarisation measurements, X and Y \citep{2017SoPh..292...87S}. 
Net linear polarization should be absent from quiet solar observation, and any such polarization in Band~3 or Band~6 should be negligible in comparison to current instrumental precision.
It is therefore possible to attribute any observed difference between the solar synthesised images of XX and YY-data to noise. 
The noise level measurement is taken as the standard deviation of a gaussian function fitted to the distribution of values in the XX minus YY image.

\cite{2017ApJ...841L...5S} quote a brightness temperature enhancement for the moving plasmoid (box 2) of 145~K with a calculated noise level for the dataset of 11~K. 
We replicate the 11~K noise level value presented by \cite{2017ApJ...841L...5S} by estimating for the full Band~3 bandwidth image synthesised over the entire observations duration. 
Our value is representative of the noise level in the images at a single time 2~s cadence observation within the particular scan of interest. 
Using this method we calculate a brightness temperature enhancement of 220~K with a calculated noise level of 14~K. 
Whilst the overall lightcurves are very similar, our calculated brightness temperature enhancement value differs somewhat from the value quoted by \cite{2017ApJ...841L...5S}.
This may be due to differences in the definition of the box dimensions and time ranges used in either study or through differences in calibration.
For example, in this study we have only used the calibration methods presented in the reference scripts for the SV data which does not contain further corrections such as self-calibration. 

We then follow the same procedure to calculate the brightness temperature at the four constituent spectral windows of Band 3; 93, 95, 105 and 107 GHz~\citep{2017SoPh..292...88W}.
The resulting brightness temperature curves can be seen in Figures~\ref{fig:box1lightcurve} and \ref{fig:box2lightcurve}. 

The noise level of each sub-band was calculated again for a single time step of 2s using the method given in \cite{2017SoPh..292...87S}.  
The gaussian fitted noise distributions can be seen in Figure \ref{fig:spw_noise}. 
The  gaussian fit to the data is noticeably better for spectral windows 0 and 1 when compared to 2 or 3. 
Analysis of the kurtosis of each dataset shows that spectral windows 2 and (in particular) 3 have non-gaussian distributions. 
The exact reason for this needs to be addressed in a future study. 

The noise levels quoted so far describe the representative value of the noise in the image, and are thus used as the detection limit of the image, and cannot be used as the error of the brightness temperature of a specified region.
We therefore follow the following procedure to calculate the brightness temperature enhancement noise at the four constituent spectral windows of Band~3 within each observational box. 
The value for the noise in each sub-band was calculated using half of the average of the absolute difference between the XX and YY data in each specified region and at each of the timesteps in the scan.
It was found that the noise evaluated in this manner was different between observational boxes but did not evolve in time, remaining at a constant value, $\sigma_{box}(\nu)$.
As the number of timesteps in both the background and enhanced phases were kept equal at $N=29$, the propagated noise for the enhancement at each sub-band for each box was calculated using the equation:
\begin{equation}\label{eq:error_propagation}
\sigma_{E,noise}(box, \nu) = \sqrt{ \sigma_{box}(\nu)^2\frac{2}{N}}
\end{equation}

\subsection{Flux scale accuracy}\label{sec:flux_scale_accuracy}

According to section 10.4.8 of the ALMA Cycle 6 Technical Handbook\footnote{\url{https://almascience.eso.org/documents-and-tools/cycle6/alma-technical-handbook}} \citep{ALMA_handbook} there is a limit to the accuracy of the flux, and thus brightness temperature, scale of an observation with ALMA.
This accuracy limit is said to increase with frequency and is quoted for ALMA Band~3 to be 5\%.
The 5\% value is a conservative estimate as the flux scale uncertainty is built on a combination of sources including; system temperature measurement, absolute flux calibration and temporal gain calibration.
Because of this the true uncertainty in the flux scale accuracy will often be less than this value. 
To model this we have assumed a normally-distributed random uncertainty where the mean is zero and $3\sigma$ is equal to the 5\% limit.
Including this scaling accuracy limit as a systematic error the standard deviation of the normally-distributed brightness temperature enhancement error becomes:
\begin{equation}\label{eq:error}
\begin{aligned}
\sigma_{E}(box,\nu)^2 = \sigma_{E,noise}(box, \nu)^2 + \\ (\frac{0.05}{3}\times T_{B,background}(box, \nu))^2 + \\ (\frac{0.05}{3}\times T_{B, enhanced}(box, \nu))^2,
\end{aligned}
\end{equation}
where $T_{B,background}(box, \nu))$, and $ T_{B,enhanced}(box, \nu))$ are the interferometric brightness temperatures of the background and enhanced phases shown in figures~\ref{fig:box1lightcurve} and~\ref{fig:box2lightcurve} for a given box and spectral window, respectively.

The resulting enhancement at each spectral window, and the standard deviation of their respective normally-distributed uncertainties are given in Table~\ref{table:results}.
\subsection{Brightness temperature enhancement spectrum}\label{sec:enhancement}

We define the brightness temperature enhancement as the difference between the brightness temperature emitted during a period of enhancement and its background value. 
Assuming an isothermal enhancing plasma the equation for the frequency dependent brightness temperature enhancement, $E({\nu})$ is;
\begin{equation}\label{eq:enhancement}
E({\nu}) = (T - T_{\mathrm{B0}}(\nu))(1 - \mathrm{e^{-\tau (\nu)}}) , 
\end{equation}
where $T$ is the  temperature of the enhancing plasma, $T_{\mathrm{B0}}(\nu)$ and $\tau(\nu)$ are the frequency-dependent background quiet Sun brightness temperature and optical thickness, respectively.
The sign of the enhancement depends on whether the temperature of the plasma is greater (positive enhancement) or less (negative enhancement) than the background brightness temperature value. 

The logarithmic-scale gradient of the enhancement spectrum (Equation~\ref{eq:enhancement}) follows a similar relation with optical thickness to the off-limb version described in Equation~\ref{eq:spectral_gradient_off_limb} but with an additional term, $\beta$, dependent on frequency and on the background solar spectrum:
\begin{equation}\label{eq:gradient}
\frac{\mathrm{d\,log}(E)}{\mathrm{d\,log}(\nu)} = \beta - \alpha \frac{2\tau_{\nu}}{\mathrm{e}^{\tau_{\nu}}-1}, 
\end{equation}
where, 
\begin{equation}\label{eq:beta}
\beta = \frac{-\frac{\mathrm{d}T_{\mathrm{B0}}}{\mathrm{d}\nu}\nu}{T - T_{\mathrm{B0}}}.
\end{equation}
Due to the structure of the solar chromosphere where the background emission is formed and the width of the observing band, the gradient of the background spectrum, $-\frac{\mathrm{d}T_{\mathrm{B0}}}{\mathrm{d}\nu}$, will be a small negative value.
$\beta$ will thus be a negative or positive factor depending on whether the constant temperature, $T$, is less than or greater than the brightness temperature of the background emission at band centre, $T_{\mathrm{B0}}$, respectively. 
The magnitude of the $\beta$ term will be mostly small except when near to the discontinuity at $T = T_{\mathrm{B0}}$. 

\begin{figure}
	\centering
	\includegraphics[width=\linewidth]{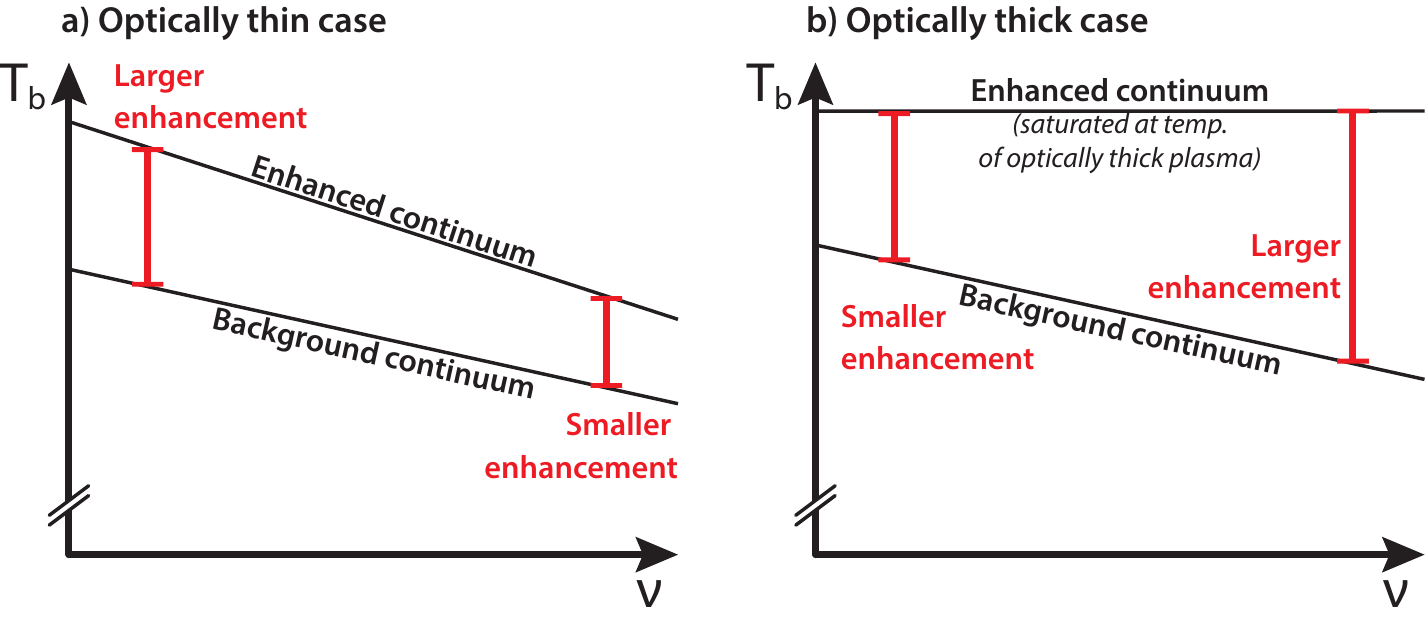}
	\caption{Schematic diagrams showing the change in brightness temperature enhancement with frequency for an optically thin or optically thick enhancing 
		isothermal material.}\label{fig:sketch}
\end{figure}

For a fully optically thin plasma, $\tau \ll 1$, the gradient of the enhancement spectrum will tend towards $\frac{\mathrm{d\,log}(E)}{\mathrm{d\,log}(\nu)} = \beta - 2\alpha$. 
For fully optically thick plasma,  $\tau \gg 1$, it shall tend towards $\frac{\mathrm{d\,log}(E)}{\mathrm{d\,log}(\nu)} = \beta$.
The reason for this transition is that optically thin source material will produce a slope dominated by the same frequency dependence as Equations~\ref{eq:opthin_isoT} and~\ref{eq:opthin_nonisoT} as there will be greater emission at lower frequencies, whilst for optically thick source material the brightness temperature at each observed frequency will reach a maximum value equal to the electron temperature of the emitting plasma.  
The quiet Sun background brightness temperature in the mm continuum decreases with increasing frequency, thus to reach the same magnitude of the electron temperature across the entire wavelength range, the enhancement spectrum would have to increase with frequency. 
There is hence a transition between a negative-gradient enhancement spectrum and a positive-gradient enhancement spectrum when the enhancing plasma's optical thickness increases significantly above unity.
A schematic graph of this mechanism is given in Figure~\ref{fig:sketch}.


\section{Results} \label{sec:results}
\begin{figure}[t]
	\centering
	\includegraphics[width=\linewidth]{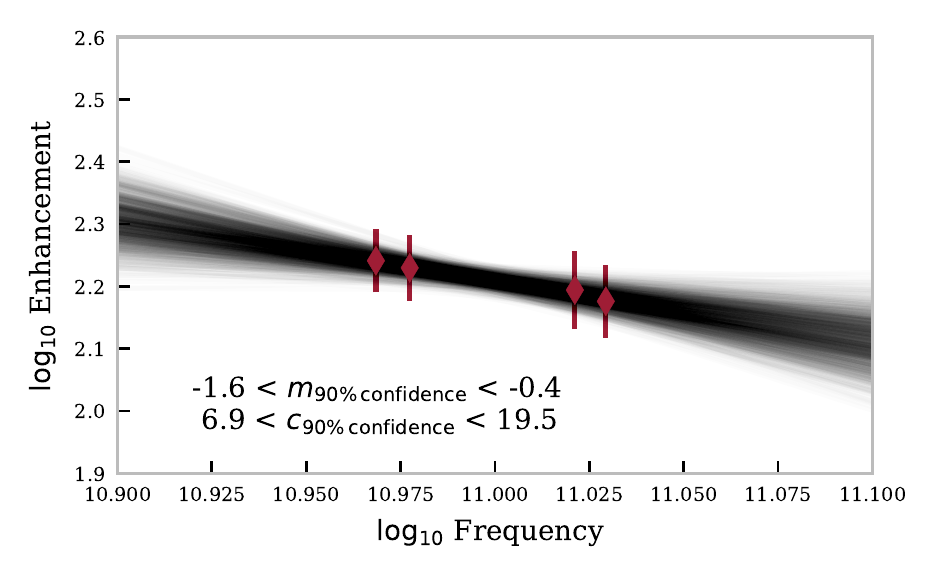}
	\caption{A subset of the MCMC fitted logarithmic-scale mean mm continuum brightness temperature enhancement spectra for the Box~1 region coinciding with the XBP from figure \ref{fig:boxes} is shown as overlaid grey lines.
		The red data points show the observed brightness temperature measurement, with the bars representing the $3\sigma$ value of the normally-distributed likelihood functions used in the statistical model. The values of $\sigma$ for these error bars are propagated in logarithmic space from the values given in Table~\ref{table:results}. 
		The range of values for the gradient and intercept of the spectral fits to 90\% confidence are shown on the plot.}\label{fig:spectrum_box1}
\end{figure}

\begin{figure}
	\centering
	\includegraphics[width=\linewidth]{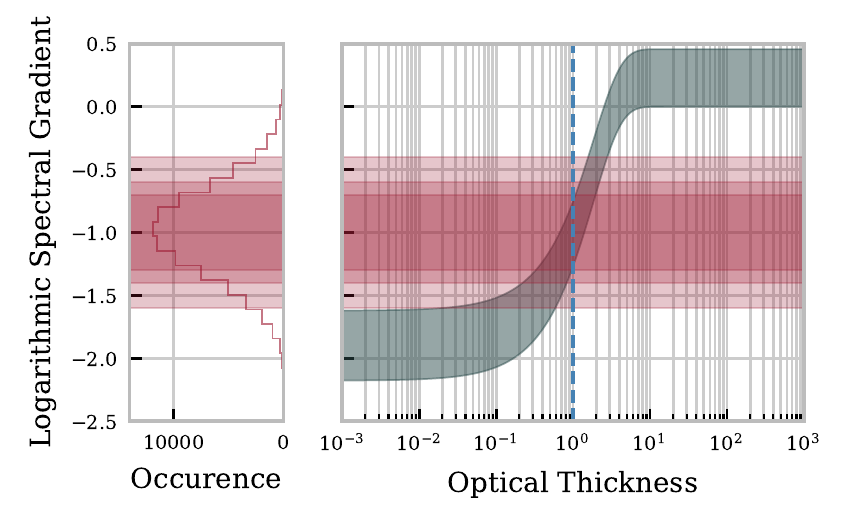}
	\caption{Graphs showing the relation between optical thickness and logarithmic-scale mm continuum spectral gradient for the structure in box~1 of Figure~\ref{fig:boxes}.
		In the left panel a histogram of the results from the MCMC sampling of our statistical model is shown.
		The regions in shades of red represent the 90, 75, and 60\% confidence intervals for the MCMC fitted observed logarithmic continuum enhancement gradient, calculated as shown in Figure~\ref{fig:spectrum_box1}.
		In the right panel the green region shows the curve of the diagnostic relationship between optical thickness and spectral gradient defined by Equation~\ref{eq:gradient} and calculated for box~1 in green.  
		The regions where the green and red colours overlap thus show the possible ranges for the optical thicknesses of the structure given the observed data and the degree of condidence in the result.
		The dashed blue line shows the location of the $\tau=1$ line.}
	\label{fig:diagnostic_box1}
\end{figure}

\begin{figure}[ht!]
	\centering
	\includegraphics[width=\linewidth]{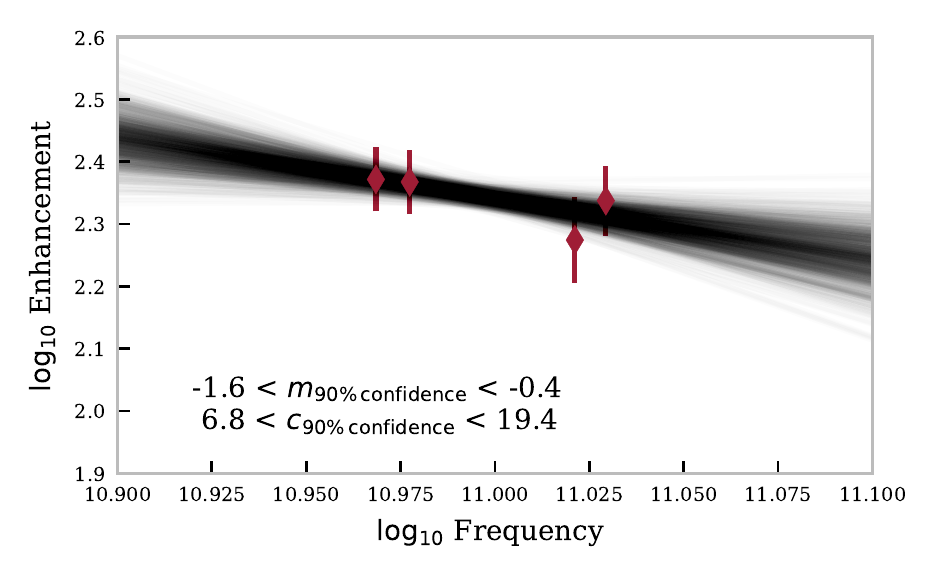}
	\caption{Same as Figure~\ref{fig:spectrum_box1} for Box~2.}\label{fig:spectrum_box2}
\end{figure}

\begin{figure}
	\centering
	\includegraphics[width=\linewidth]{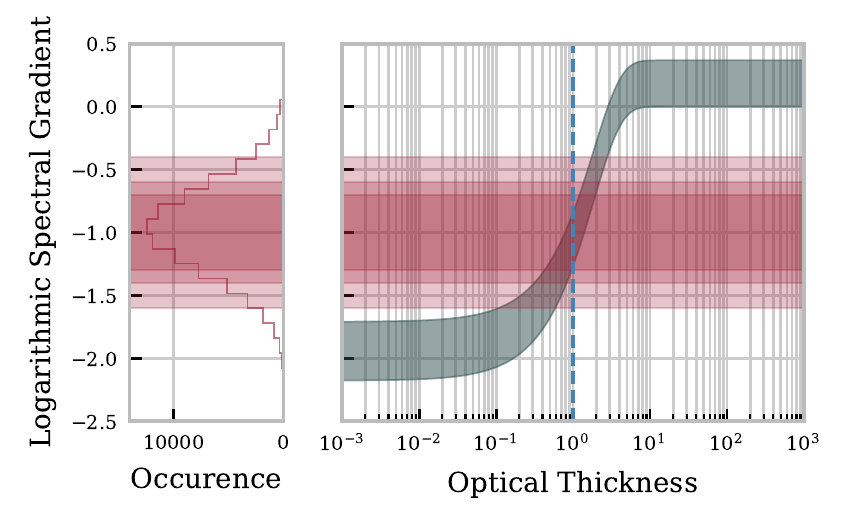}
	\caption{Same as Figure~\ref{fig:diagnostic_box1} for Box~2.}\label{fig:diagnostic_box2}
\end{figure}

Synthesised images were produced using the method described in Section \ref{sec:obs}. 
The inferred brightness temperature enhancement of the stationary XBP-associated enhancement observed in box~1 and the moving plasmoid ejection observed in box~2 using the four constituent sub bands of ALMA Band~3 are presented in Table~\ref{table:results}. 

\subsection{Box~1: Stationary enhancement coincident with XBP}\label{sec:box1}

The logarithmic-scale millimeter continuum enhancement spectrum for the stationary enhancement observed coincident with the XBP in box 1 of figure~\ref{fig:boxes} is shown in figure \ref{fig:spectrum_box1}.
As the separation in frequency across Band 3 is relatively small we assume that the curve of the mm continuum spectrum can be approximated with a straight line; $\mathrm{log_{10}(}E) = m\mathrm{log_{10}}(\nu) +c$, where $m$ is the gradient and $c$ is the y-intercept, regardless of the optical thickness regime. 
To fit the enhancement spectrum we have decided to use a bayesian linear regression method as to make best use of the uncertainty distributions defined in Sections~\ref{sec:noise} and \ref{sec:flux_scale_accuracy}.  
Another advantage to this method is that we can produce results in the form of a posterior probability distribution.
The statistical model was created by defining suitable, logarithmic likelihood and prior distributions. 
The likelihood function was defined to be a normal distribution with standard deviation equal to the values quoted in Table~\ref{table:results} propagated into logarithmic space. 
The prior distributions for the two desired fitting parameters, i.e spectral gradient and y-intercept, have been set as non-informative uniform distributions. 
The width of each uniform distribution was set to be wide enough to encompass all possible values for the parameters were the inference carried out with a less-informed least-squares method.
With the model defined, we sampled it using a python implementation of the affine-invariant ensemble sampler for Markov-chain Monte-Carlo (MCMC) method (EMCEE) \citep{2013PASP..125..306F}. 
We sampled using 100 chains, each with 1000 steps including tuning.
A subset of the sampled fitting models is shown with the observed sub-band data in Figure~\ref{fig:spectrum_box1}.
The simplest deduction from the spectrum is that as the enhancement is positive the electron temperature of the plasma must be greater than the brightness temperature of the background atmosphere. 
From the posterior distribution, we find within the 90\% confidence intervals, that the spectral gradient ranges between -1.6 and -0.4  which signifies that the optical thickness of the plasma is likely to be within the transition between fully optically thin and optically thick material, as discussed in Section~\ref{sec:enhancement} and Figure~\ref{fig:sketch}. 
The confidence regions are estimated using the percentile method.
Due to the finite number of samples during the MCMC process the estimated confidence intervals can be subject to some small variation between different runs (i.e. of order $\sim10^{-2}$).
To account for this potential variation in these estimated values we round them to one decimal places for use in all further calculations.
To determine an estimate for the optical thickness of the enhancing plasma we defined the relevant diagnostic curve for the observation through Equation~\ref{eq:gradient}. 

The $\alpha$ correcting factor, due to a non-zero rate of change of the gaunt factor across the frequency-band, defined in Equation~\ref{eq:alpha}, was estimated by calculating the gaunt factor and the rate of change of the gaunt factor with frequency for ALMA Band~3 and a wide range of potential constant temperatures ($T = 10^3 - 10^6$~K). 
The values for the gaunt factor were interpolated from the table of calculated values from \cite{2014MNRAS.444..420V}.
Assuming that all potential constant temperatures are equally likely the minimum and maximum values for $\alpha$ are 1.04 and 1.09, respectively \citep{Rodger&Labrosse2018}.

The other factor necessary to estimate for the diagnostic curve is $\beta$ (Equation~\ref{eq:beta}).
Estimating $\beta$ requires an estimate of the gradient of the background brightness temperature spectrum, which is defined by the temperature structure of the solar chromosphere and the width of ALMA Band~3. 
To estimate this value we have adopted the quiet-Sun model C7 from \cite{2008ApJS..175..229A} to give an example continuum spectrum for Band~3.
We assume a purely hydrogen plasma and a solely thermal bremsstrahlung emission mechanism.
The absorption coefficient for thermal bremsstrahlung is calculated as described in \cite{1985ARA&A..23..169D};
\begin{equation}
\kappa_{\nu} = 1.77\times10^{-2}\frac{n_e}{\nu^2T^{\frac{3}{2}}}\sum_iZ_i^2n_ig_{\mathrm{ff}},
\end{equation}
with again the free-free gaunt factor, $g_{\mathrm{ff}}$, as interpolated from the table of calculated values from \cite{2014MNRAS.444..420V}.
From the calculated absorption coefficient and temperature values of the C7 model we integrate the equation;
\begin{equation}
T_{\mathrm{B}} = \int \kappa_{\nu} T \mathrm{e}^{-\tau_{\nu}} \mathrm{d}s, 
\end{equation} 
along the path, $s$, to find the background brightness temperature values for the atmosphere. 
This method is similar to \cite{2012SoPh..277..31H, 2017A&A..605A.125S}.
The value for the background spectral gradient for C7 and ALMA Band~3 we find is $\sim$-9$\times10^{-10}$.
From Equation~\ref{eq:enhancement} it can be seen that $(T-T_{\mathrm{B0}}(\nu)) \geq E(\nu)$ must be true, because of this we only evaluate the $\beta$ term between the values $E(\nu) \leq (T-T_{\mathrm{B0}}(\nu)) \leq 10^6$, where we use the full-band ALMA Band~3 enhancement value for box~1 (Table~\ref{table:results}) as $E(\nu)$.
Restricting the range of $(T-T_{\mathrm{B0}}(\nu))$ values considered like this allows us to avoid the discontinuity found in Equation~\ref{eq:beta}.
Following this procedure the minimum and maximum values for $\beta$ are $\sim$0.00 and 0.46, respectively.

The diagnostic curve for the optical thickness of the enhancing plasma is thus made using the maximum and minimum values for the two factors; $\alpha$ and $\beta$.
Through plotting this curve and the confidence intervals of the fitted enhancement gradient it is possible to estimate the optical thickness/optical thickness regime of the plasma through the positions where the regions intersect. 

The figure showing this method is presented in figure~\ref{fig:diagnostic_box1}.
From this it can be seen that we can make an inference on the maximum, and the minimum optical thickness of the stationary enhancement, coincident with an XBP.
Within 90\% confidence, the result we find  is that 0.02  $\leq \tau_{90\%\,\mathrm{confidence}} \leq$ 2.78.

Although part of the estimated optical thickness of the plasma in box~1, within 90\% confidence, lies above unity it is not high enough to be in the regime where the brightness temperature may be used as a direct analogue of the electron temperature. 
We can, however, estimate the  difference between the electron temperature of the plasma and the background brightness temperature ($T-T_{\mathrm{B0}}$) using the estimated optical thickness and Equation~\ref{eq:enhancement}.
In this manner we estimate the value to be 170 $\leq T-T_{\mathrm{B0}}\leq$ 7900~K.
If we were to assume the background brightness temperature for Band~3 emission to be a typical value of $\approx7300\pm100$~K \citep{2017SoPh..292...88W} we would thus expect the electron temperature in box~1 to be between $\approx$ 7370 -- 15300~K.
This assumption was checked by viewing the ALMA single-dish images during the observation which allowed us to conclude that the \cite{2017SoPh..292...88W} quoted value for the typical mm background value in Band~3 is an applicable assumption for this study.
If the plasma had the maximum or minimum possible optical thicknesses, as measured using our method, we would expect it to have a maximum emission measure of $\sim$ 0.06 -- 3 $\times 10^{29}\,\mathrm{cm^{-5}}$, following Equation~\ref{eq:opthin_isoT}.

\subsection{Box~2: Moving enhancement from plasmoid ejection}

We analysed the moving enhancement due to the plasmoid ejection in box 2 in the same manner as box 1 as outlined in section~\ref{sec:box1}. 
A resulting subset of the MCMC fitted continuum brightness temperature enhancement spectra, and the 90\% confidence intervals for the two fitting parameters for box~2, can be seen in figure~\ref{fig:spectrum_box2}.
Again the first noticeable diagnostic indications are that the enhancement is positive and the gradient of the spectrum is negative, meaning that the temperature of the structure must be greater than the background brightness temperature value and that the plasma is either optically thin or near the transition to optically thick. 

In creating the optical thickness diagnostic curve (Equation~\ref{eq:gradient}) we follow the same procedure for box~2 as described in Section~\ref{sec:box1}.
We use the same bounds for the $\alpha$ factor here as for box~1, whilst we calculate slightly different bounds for the $\beta$ factor at $\sim$0.00 -- 0.37, due to the different value for the enhancement at Band~3 band-centre.
The plot showing the diagnostic curve compared to the 90\% confidence interval estimates of the  mm enhancement spectral gradient for box~2 defining the moving plasmoid enhancement observation is shown in figure~\ref{fig:diagnostic_box2}.

In the same manner as the previous analysis the region in figure~\ref{fig:diagnostic_box2} where the observed gradient and diagnostic curve overlap shows the range of possible optical thicknesses for the plasmoid. 
From this figure it can be seen that the optical thickness of the plasmoid ranges from 0.11 -- 2.78, lying in the transition region between optically thin and optically thick material.
From the enhancement at band centre and this optical thickness estimation, the difference between the temperature of the plasmoid and the background brightness temperature ($T-T_{\mathrm{B0}}$) is calculated to be 240 -- 2160~K.

Assuming again a typical background quiet Sun brightness temperature at Band~3 of 7300$\pm100$~K \citep{2017SoPh..292...88W} would give an electron temperature of the plasmoid of $\sim$7440 -- 9660~K.
Using this temperature estimation and the estimated optical thickness we find the maximum emission measure of the moving plasmoid structure to range between $\sim$0.2 -- 3$\times 10^{29}\,\mathrm{cm^{-5}}$.
Assuming that the width of the plasmoid is equal to its extent on the disc ($\sim4\arcsec\approx3000$~km) the electron density of the plasma would be in the range $\approx$ 0.7 -- 3$\times 10^{10} \,\mathrm{cm^{-3}}$.

In \cite{2017ApJ...841L...5S} the authors conclude that the moving plasmoid is roughly consistent with either an isothermal $\approx$ $10^5$~K plasma that is optically thin at 100~GHz (density of $\approx10^9$~$\mathrm{cm^{-3}}$), or a cool optically thick plasma core of temperature $\approx10^4$~K and density $\geq2\times10^{10}\,\mathrm{cm^{-3}}$.
The results from our study support more closely the \cite{2017ApJ...841L...5S} case where the plasmoid is cool and optically thick, however, the estimated optical thickness in this study lies across the transition from optically thin to optically thick material. 

\section{Discussion}\label{sec:discussion}

Whilst the equations used in the analysis for this study have been derived from an isothermal assumption it is possible that the objects observed in boxes~1 and 2 are in some way multi-thermal.
It has been found, however, in \cite{Rodger&Labrosse2018}, that the isothermal assumption in the relationship between logarithmic spectral gradient and optical thickness holds well for a multi-thermal plasma for optical thickness $\leq 1$.
Beyond $\tau = 1$ the logarithmic spectral gradient relationship with optical thickness is expected to deviate from the isothermal case increasingly with increasing optical thickness.
The estimated optical thickness for a multi-thermal plasma passed the $\tau=1$ line could be expected to be under-estimated compared to its true value. 
In both observational boxes for this study we have found optical thicknesses close to the $\tau=1$ line, where the expected relationship derived under the isothermal assumption should still mostly agree with a multi-thermal case. 

A source of uncertainty not considered within our estimation of the optical thickness is the uncertainty in the gradient of the background brightness temperature spectrum, which is necessary for the calculation of the $\beta$ factor in Equation~\ref{eq:gradient}. 
In this study we have used a value calculated from the atmospheric Quiet-Sun model C7 of \cite{2008ApJS..175..229A}.
In future studies, when the uncertainties on absolute brightness temperatures are better understood, it may be beneficial to use observed spectral gradient values taken from the feathered total-power and interferometric ALMA data. 
In the estimation of the emission measure and the temperature of the structures, another source uncertainty could originate from assuming the typical ALMA Band~3 background brightness temperature of 7300$\pm100$ suggested by \cite{2017SoPh..292...88W}. 
Again, in the future, this shall be addressed through the use of absolute brightness temperature observations.

The largest source of uncertainty in the data is due to the accuracy of the flux scale determination. 
If this source of uncertainty would become smaller or better understood in future ALMA cycles this would improve the quality of this diagnostic method. 
This source of uncertainty also increases with increasing frequency, such that, once lower frequency ALMA Bands, such as Bands~1 and 2, become available to solar observations they may provide an improved wavelength range for this technique.
Future efforts to determine the slope of the logarithmic millimetre continuum could also be better understood through the addition of more, and in particular more spread out in frequency, brightness temperature measurements.

\section{Conclusions} \label{sec:conc}

This study provides the first sub-band spectral analysis of an ALMA solar observation. 
Sub-band analysis of the logarithmic mm continuum brightness temperature spectrum proves to be a potentially powerful technique for diagnosing plasma optical thickness and thus other plasma parameters such as electron temperature and emission measure, provided that suitable uncertainties are defined. 
We have shown this for the first time through the calculation of the logarithmic mean brightness temperature enhancement spectrum across the four sub-bands of ALMA Band 3 in two regions associated with an X-ray Bright Point (XBP) and plasmoid ejection event of 17th of December 2015.
Using a bayesian linear regression method we found the posterior probability distributions for resulting straight line trends. 
The 90\% confidence regions for the gradient of the spectra were compared to the expected optical thickness versus spectral gradient diagnostic curve for an ALMA Band~3 observation of an on-disc structure of given band-centre brightness temperature enhancement, finding the possible optical thicknesses where the two regions overlapped. 
From this analysis we show that the optical thickness of the stationary enhancement is between 0.02 $\leq \tau \leq$ 2.78, whilst the moving enhancement has 0.11 $\leq \tau \leq$ 2.78, where both lie entirely in the transition region between optically thin and optically thick plasma. 
Assuming a typical Quiet Sun background brightness temperature of 7300$\pm100$~K \citep{2017SoPh..292...88W} we expect an electron temperature for the stationary enhancement of $\approx$ 7370 -- 15300~K and between 7440 -- 9660~K for the moving plasmoid enhancement.
Although the analysis presented here for the moving plasmoid feature suggests a material with optical thickness near to the transition between optically thin and thick material, it supports better the case presented by \cite{2017ApJ...841L...5S} where the plasmoid has a cool core of temperature $\approx10^4$~K plasma with density of $\geq2\times10^{10}$~$\mathrm{cm^{-3}}$ against the option of a fully optically thin plasmoid with a temperature of $\approx$ $10^5$~K and a density of $\approx10^9\,\mathrm{cm^{-3}}$.

\section{Acknowledgements}
ASR acknowledges support from the STFC studentship ST/N504075/1. 
NL and LF acknowledge support from STFC grant ST/P000533/1. 
SW and MS are supported by the SolarALMA project, which has received funding from the European Research Council (ERC) under the European Union’s Horizon 2020 research and innovation programme (grant agreement No. 682462), and by the Research Council of Norway through its Centres of Excellence scheme, project number 262622.
PJAS acknowledges support from the University of Glasgow's Lord Kelvin Adam Smith Leadership Fellowship.
ASR would like to acknowledge the opportunity for learning and collaboration presented by an STFC-funded long-term attachment to the Institute of Theoretical Astrophysics at the University of Oslo, where much of the groundwork to this study took place. 
The authors would like to thank the referee for their insight and help in the process of producing this article.
The authors would also like to thank Hugh Hudson and Daniel Williams for their constructive comments and discussion. 

This paper makes use of the following ALMA data: ADS/JAO.ALMA\#2011.0.00020.SV. ALMA is a partnership of ESO (representing its member states), NSF (USA) and NINS (Japan), together with NRC (Canada) and NSC and ASIAA (Taiwan), and KASI (Republic of Korea), in cooperation with the Republic of Chile. The Joint ALMA Observatory is operated by ESO, AUI/NRAO and NAOJ.

\bibliography{paper}




\end{document}